\documentclass{PoS}

\usepackage{amsmath,amsfonts,amssymb,booktabs,mathtools}
\title{Using a new analysis method to extract excited states in the scalar meson sector}

\ShortTitle{Using a new analysis method to extract excited states in the scalar meson sector}
        
\author{\speaker{Jacob Finkenrath}${}^{1}$, Constantia Alexandrou${}^{1,2}$, Joshua Berlin${}^{3}$, Mattia Dalla Brida${}^{4,5}$, Theodoros Leontiou${}^{6}$, Marc Wagner${}^{3}$\\

\email{alexand@ucy.ac.cy},
\email{berlin@th.physik.uni-frankfurt.de},
\email{mattia.dalla.brida@desy.de},
\email{j.finkenrath@cyi.ac.cy},
\email{t.leontiou@frederick.ac.cy},
\email{mwagner@th.physik.uni-frankfurt.de}\\

      \begin{flushleft}
        ${}^{1}$Computation-based Science and Technology Research Center, The Cyprus Institute, \\ 20 Kavafi Street, 2121 Nicosia, Cyprus\\
        ${}^{2}$Department of Physics, University of Cyprus, P.O. Box 20537, 1678 Nicosia, Cyprus\\
        ${}^{3}$Goethe-Universit\"at Frankfurt am Main, Institut f\"ur Theoretische Physik, \\  Max-von-Laue-Stra{\ss}e 1, D-60438 Frankfurt am Main, Germany\\ 
        ${}^{4}$NIC, DESY, Platanenallee 6, 15738 Zeuthen, Germany\\
        ${}^{5}$Dipartimento di Fisica, Universit\`a di Milano-Bicocca, and INFN, \\
                 sezione di Milano-Bicocca, Piazza della Scienza 3, I-20126 Milano, Italy \\
        ${}^{6}$Department of Mechanical Engineering, Frederick University, 7 Y. Frederickou Str., \\ Pallouriotisa, Nicosia 1036, Cyprus
      \end{flushleft}
      }        

\abstract{We explore a method to extract energy eigenstates, called Athens Model Independent Analysis Scheme (AMIAS),
which is an alternative to solving standard Generalized Eigenvalue Problems (GEVP). The method is based on statistically
sampling the space of fit parameters according to the $\chi^2$ value of the fit function. The method is particularly 
suited for correlators or correlation matrices with strong contributions from several energy eigenstates and for rather 
noisy data, e.g.\ for correlators with disconnected and partly disconnected diagrams. We apply the method to the analysis 
of the $J^P = 0^+$ channel in the context of our investigation of the $a_0(980)$ meson and point out advantages compared to the GEVP.
}

\FullConference{The 34nd International Symposium on Lattice Field Theory\\
                 25-30 July, 2016\\
                 University of Southampton, UK}

\newcommand{\gtapprox}{\raisebox{-0.5ex}{$\,\stackrel{>}{\scriptstyle\sim}\,$}}
\newcommand{\ltapprox}{\raisebox{-0.5ex}{$\,\stackrel{<}{\scriptstyle\sim}\,$}}

\begin{document}


\section{Introduction}
\label{sec:mot}

In this work we explore AMIAS, a rather new analysis method to extract energy differences and
amplitudes from lattice QCD correlators and correlation matrices. The basic idea is to consider
a very large number of fits to these correlators using Monte Carlo methods, which will lead to 
probability distributions for the fit parameters, that are the energy differences and amplitudes.
This avoids the necessity of identifying plateau regions for effective energies. In particular, for unstable 
systems and resonances identifying plateaus can be very challenging, since the signal-to-noise ratio
is typically rather poor for large temporal separations. Moreover it overcomes the limitation of extracting at most
$N$ energy eigenstates, when $N$ interpolating fields are used (as it is e.g.\ the case, when doing a standard GEVP analysis)

We apply AMIAS to analyze correlators computed for studying the $a_0(980)$ meson, which has quantum
numbers $I(J^P) = 1(0^+)$ and mass $m_{a_0(980)} \approx 980 \, \textrm{MeV}$ \cite{Agashe:2014kda}. We
use interpolating fields with both two-quark and a four-quark structures formed by an up-quark $u$ and
an anti-down quark $\bar{d}$ and except for one case a strange and an anti-strange quark $s$  and $\bar{s}$. The four-quarks
are arranged as a meson-meson interpolating field or in a diquark-anti-diquark combination,
which can probe different possible tetraquark structures of $a_0(980)$. We consider six interpolating fields,
\begin{eqnarray}
\label{EQN002} & & \hspace{-0.7cm} \mathcal{O}^1 \coloneqq \mathcal{O}^{q \bar{q}} = \frac{1}{\sqrt{V_s}} \sum_{\bf{x}} \Big({\bar d}({\bf x}) u({\bf x})\Big) \\
\label{EQN007} & & \hspace{-0.7cm} \mathcal{O}^2 \coloneqq \mathcal{O}^{K \bar{K}, \ \textrm{point}} = \frac{1}{\sqrt{V_s}} \sum_{\bf{x}} \Big({\bar s}({\bf x}) \gamma_5 u({\bf x})\Big) \Big({\bar d}({\bf x}) \gamma_5 s({\bf x})\Big) \\
 & & \hspace{-0.7cm} \mathcal{O}^3 \coloneqq \mathcal{O}^{\eta_{s} \pi, \ \textrm{point}} = \frac{1}{\sqrt{V_s}} \sum_{\bf{x}} \Big({\bar s}({\bf x}) \gamma_5 s({\bf x})\Big) \Big({\bar d}({\bf x}) \gamma_5 u({\bf x})\Big) \\
 & & \hspace{-0.7cm} \mathcal{O}^4 \coloneqq \mathcal{O}^{Q \bar{Q}} = \frac{1}{\sqrt{V_s}} \sum_{\bf{x}} \epsilon_{a b c} \Big({\bar s}_b({\bf x}) (C \gamma_5) {\bar d}_c^T({\bf x})\Big) \epsilon_{a d e} \Big(u_d^T({\bf x}) (C \gamma_5) s_e({\bf x})\Big) \\
 & & \hspace{-0.7cm} \mathcal{O}^5 \coloneqq \mathcal{O}^{K \bar{K}, \ \textrm{2part}} = \frac{1}{V_s} \sum_{{\bf x},{\bf y}} \Big({\bar s}({\bf x}) \gamma_5 u({\bf x})\Big) \Big({\bar d}({\bf y}) \gamma_5 s({\bf y})\Big) \\
\label{EQN003} & & \hspace{-0.7cm} \mathcal{O}^6 \coloneqq \mathcal{O}^{\eta_{s} \pi, \ \textrm{2part}} = \frac{1}{V_s} \sum_{{\bf x},{\bf y}} \Big({\bar s}({\bf x}) \gamma_5 s({\bf x})\Big) \Big({\bar d}({\bf y}) \gamma_5 u({\bf y})\Big) ,
\end{eqnarray}
where $V_s$ denotes the spatial lattice volume and $C$ the charge conjugation matrix. All of them couple to $a_0(980)$
and other states with the same quantum numbers. For example, the interpolating fields $\mathcal{O}_5$ and $\mathcal{O}_6$ 
mainly generate the two-meson states $K + \overline{K}$ and  $\pi + \eta$, respectively, which 
are expected to have masses close that of the $a_0(980)$. Note that the interpolating fields $\mathcal{O}_2$
and $\mathcal{O}_3$ represent two mesons located at the same point in space with only total momentum zero, but
arbitrary relative momenta involved (a structure resembling a 4-quark bound state). In contrast to that
$\mathcal{O}_5$ and $\mathcal{O}_6$ correspond to two mesons with both total and relative momentum zero.

We compute the full six by six correlation matrix including both connected and disuconnected contributions,
which we neglect in our previous studies~\cite{Abdel-Rehim:2014zwa,Berlin:2015faa,Berlin:2016zci}.
Moreover here we have increased statistical accuracy of the correlators
$C_{j k}(t) = \langle \mathcal{O}^j(t) \mathcal{O}^{k \dagger}(0) \rangle$,
included propagation of strange quarks within a timeslice and analyzed the correlators with AMIAS. 
We use an ensemble of around 500 gauge link configurations generated with 2+1 dynamical Wilson clover 
quarks and Iwasaki gauge action generated by the PACS-CS Collaboration \cite{Aoki:2008sm}. The lattice size 
is $64 \times 32^3$ with lattice spacing $a \approx 0.09 \, \textrm{fm}$ and pion mass $m_\pi \approx 300 \, \textrm{MeV}$.


\section{Correlators on a periodic lattice}
\label{sec:euccor}

A correlator on a periodic lattice with time extent $T$ can be expanded according to
\begin{equation}
C_{j k}(t) = \Big\langle \mathcal{O}^j(t) \mathcal{O}^{k \dagger}(0) \Big\rangle = \frac{\sum_{m,n} \textrm{exp}\{-\mathcal{E}_m (T-t)\} \langle m | \mathcal{O}^j | n \rangle \textrm{exp}\{-\mathcal{E}_n t\} \langle n | \mathcal{O}^{k \dagger} | m \rangle}{\sum_m \exp\{-\mathcal{E}_m T\}}
\end{equation}
with energy eigenstates $| m \rangle$ and corresponding energy eigenvalues 
$\mathcal{E}_0 \leq \mathcal{E}_1 \leq \mathcal{E}_2 \ldots$ ($| 0 \rangle = | \Omega \rangle$ denotes the vaccum). 
This correlator can also be expressed in a more convenient form,
\begin{equation}
C_{j k}(t) = \Big\langle \mathcal{O}^j(t) \mathcal{O}^{k \dagger}(0) \Big\rangle = \frac{\sum_{m,n} c^j_{m,n} (c^k_{m,n})^\ast \textrm{exp}\{-(\mathcal{E}_m + \mathcal{E}_n) (T/2)\} \textrm{cosh}\{\Delta \mathcal{E}_{n,m} (t - T/2)\}}{\sum_m \exp\{-\mathcal{E}_m T\}} \label{eq:cosh}
\end{equation}
with $c_{m,n}^j = \langle m | \mathcal{O}^j | n \rangle$ and $\Delta \mathcal{E}_{n,m} = \mathcal{E}_n - \mathcal{E}_m$. For 
example, if $j = k$, if $\mathcal{O}^j$ probes the sector, which contains energy eigenstate $| 1 \rangle$, if the quantum 
numbers of $| 1 \rangle$ are different from those of the vacuum $| \Omega \rangle$, and if the correlator is not contaminated
by multi-particle states (see discussion below), (\ref{eq:cosh}) reduces to
\begin{equation}
\label{EQN001} C_{j j}(t) = \Big\langle \mathcal{O}^j(t) \mathcal{O}^{k \dagger}(0) \Big\rangle \approx 2 \Big|c^j_{0,1}\Big|^2 \exp\{-\Delta \mathcal{E}_{1,\Omega} (T/2)\} \textrm{cosh}\{\Delta \mathcal{E}_{1,\Omega} (t - T/2)\}
\end{equation}
for sufficiently large $t$ and $T$. A standard technique to determine the energy difference $\Delta \mathcal{E}_{1,\Omega} = E_1 - E_\Omega$, 
the mass of state $| 1 \rangle$, from the asymptotic $t$ behavior of $C_{j j}(t)$ (Eq.~(\ref{EQN001})) is to fit
\begin{equation}
C_{j j}(t) = A \textrm{cosh}\{-\Delta \mathcal{E}_{1,\Omega} (t - T/2)\}
\end{equation}
to the lattice QCD results for $C_{j j}(t)$ with fitting parameters $\Delta \mathcal{E}_{1,\Omega}$ and $A$. Alternatively, one can also solve the equation
\begin{equation}
\frac{C_{j j}(t)}{C_{j j}(t-a)} = \frac{\textrm{cosh}\{E_\textrm{eff}(t) (t - T/2)\}}{\textrm{cosh}\{E_\textrm{eff}(t) (t-a - T/2)\}} \label{eq:effEn}
\end{equation}
with respect to $E_\textrm{eff}(t)$, where $E_\textrm{eff}(t) \approx \Delta \mathcal{E}_{1,\Omega}$. In other words, a plateau-like 
behavior of $E_\textrm{eff}(t)$ indicates the mass $\mathcal{E}_{1,\Omega}$. In practice, however, the temporal
extent $T$ of the lattice is limited and the effective energy $E_\textrm{eff}(t)$ is often very noisy for large $t$, 
rendering a reliable extraction of $\Delta \mathcal{E}_{1,\Omega}$ difficult.

\begin{figure}[htb]
\begin{center}
\begin{minipage}[c]{.45\textwidth}
  \begin{center}
  \includegraphics*[angle=0,width=\textwidth]{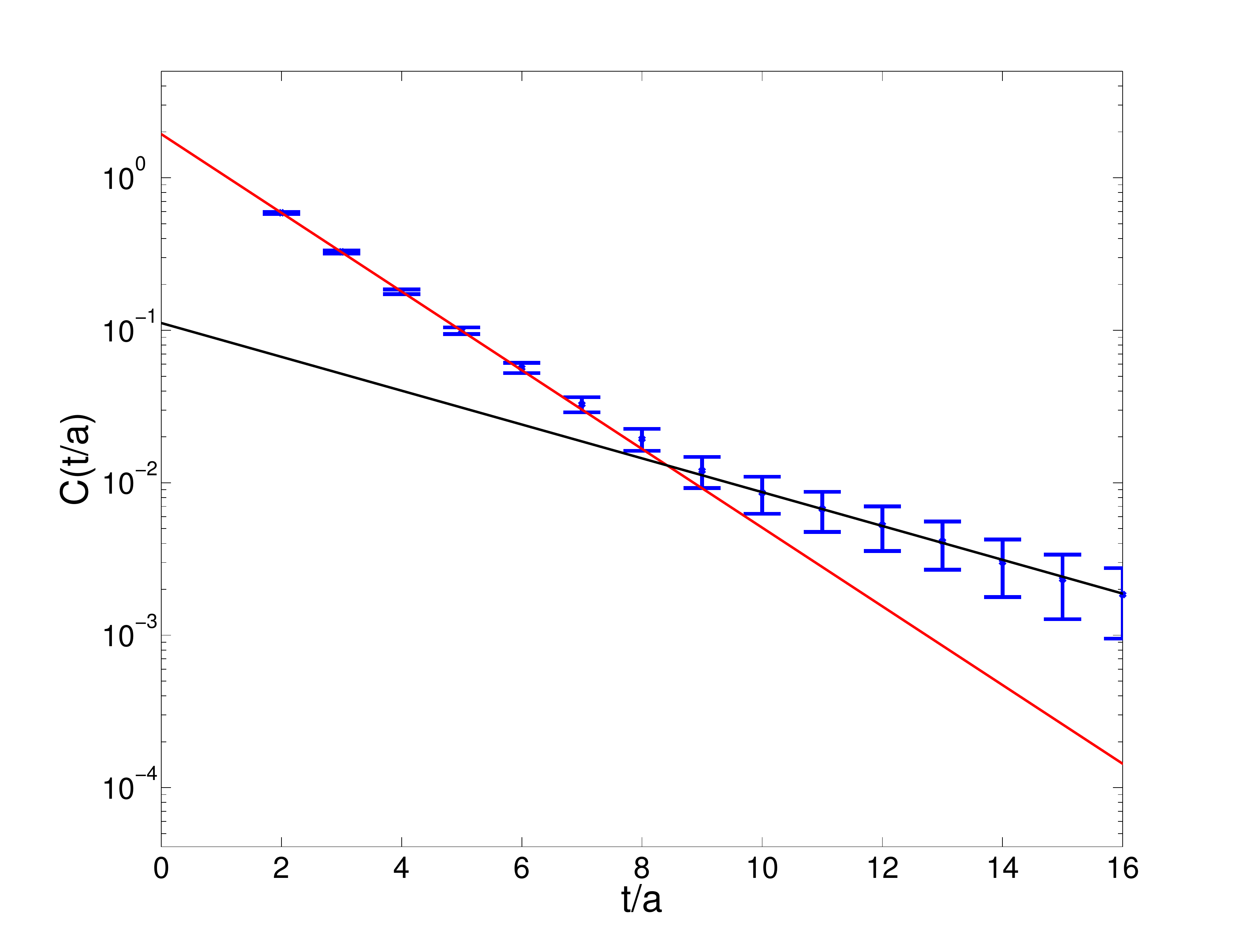} 
  \end{center}
\end{minipage}
\begin{minipage}[c]{.45\textwidth}
  \begin{center}
  \includegraphics*[angle=0,width=\textwidth]{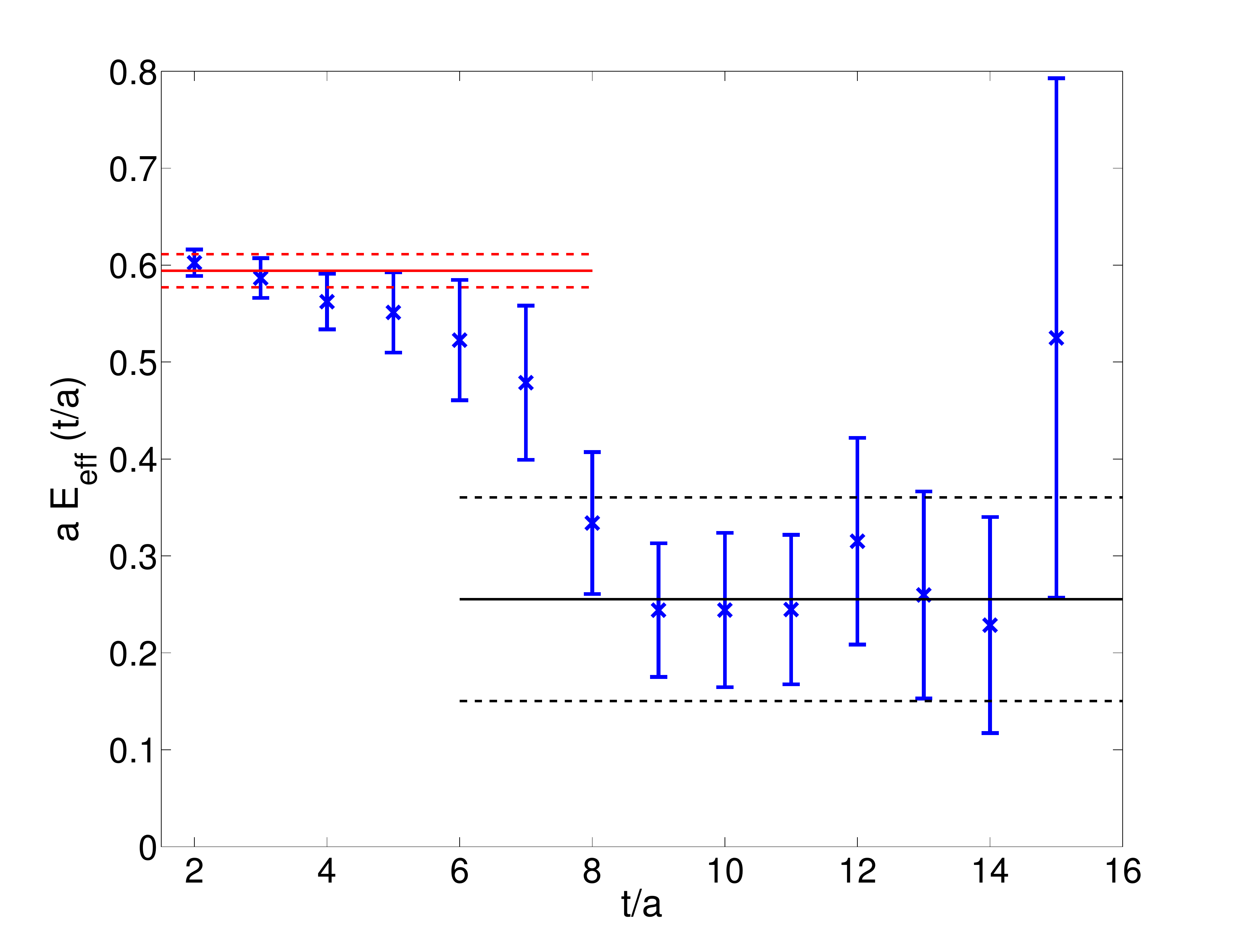}
  \end{center}
\end{minipage}
\vspace{-0.2cm}
\caption{Left: Correlator using the interpolating field $\mathcal{O}^{q \bar{q}}$. Right: Corresponding effective mass.}
\label{fig:qqbarcor}
\end{center}
\end{figure}

When multi-particle states are present in the investigated sector, the determination of low-lying masses is even more difficult.
We sketch this for a simple non-interacting two-meson system, which is generated by an interpolating field $\mathcal{O} = \mathcal{O}^{(1)} \otimes  \mathcal{O}^{(2)}$. Energy eigenstates of such a system can be written as $| n \rangle = | n_1 \rangle^{(1)} \otimes  | n_2 \rangle^{(2)}$, where $| n_1 \rangle^{(1)}$ and $| n_2 \rangle^{(2)}$ ($n_1 , n_2 = 0,1,2,\ldots$) are eigenstates of the two Hamiltonians describing the individual mesons. Energy eigenvalues of this two-meson system are $\mathcal{E}_{n_1,n_2}^{(1+2)} = \mathcal{E}_{n_1}^{(1)} + \mathcal{E}_{n_2}^{(2)}$, in particular the lowest energy eigenvalue is $\mathcal{E}_{0,0}^{(1+2)} = \mathcal{E}_{0}^{(1)} + \mathcal{E}_{0}^{(2)}$. Note that in the expansion (\ref{eq:cosh}) there is a term, which is proportional to $\cosh\{(\mathcal{E}_0^{(1)} - \mathcal{E}_0^{(2)}) (t - T/2)\}$, i.e.\ only decaying with the mass difference of the two mesons. In the region of $t \approx T/2$ this unwanted term can be more dominant than the signal term, which is proportional to $\cosh\{\mathcal{E}_{0,0}^{(1+2)} - \mathcal{E}_\Omega) (t - T/2)\}$. Thus, one has to be rather careful, when analyzing correlators of sectors containing multi-particle states. This is illustrated by Fig.~\ref{fig:qqbarcor},
where we show the correlator using the interpolating field $\mathcal{O}^{q \bar{q}}$ (Eq.\ (\ref{EQN002})). Clearly, 
there are two effective mass plateaus, $a E_{\textrm{eff}}(t) \approx 0.60$ for $t/a \ltapprox 7$, corresponding to 
the mass of a $\pi + \eta$ or a $K + \overline{K}$ two-meson state, and $a E_{\textrm{eff}}(t) \approx 0.25$ for $t/a \gtapprox 8$,
which does not correspond to the mass of any state, but to the mass difference $m_\eta - m_\pi$ of two single-meson states.

\vspace{-0.3cm}

\vspace{-1cm}
\section{AMIAS}
\label{sec:amias}

Lattice QCD results for the correlators $C_{j k}(t) = \langle \mathcal{O}^j(t) \mathcal{O}^{k \dagger}(0) \rangle$ 
with $\mathcal{O}^j$, $j = 1,\ldots,6$ defined in Eqs.~(\ref{EQN002}) to (\ref{EQN003}) can be parameterized and fitted
the expression given in Eq.~(\ref{eq:cosh}). Since statistical accuracy is limited, it is sufficient to consider a rather small number 
of energy eigenstates, i.e.\ $\sum_{m,n} , \sum_m \rightarrow \sum_{m,n = 0}^{N-1} , \sum_{m = 0}^{N-1}$ with $N \ltapprox 10$.

With AMIAS \cite{Alexandrou:2008bp,Papanicolas:2012sb,Alexandrou:2014mka} one can
determine probability distribution functions (PDFs) for the fit parameters $\Delta \mathcal{E}_{n,m}$ (energy differences)
and $c^j_{m,n}$ (amplitudes). AMIAS is able to deal with a rather large number of parameters by using Monte Carlo
techniques. In contrast to e.g.\ the GEVP, it is not necessary to identify plateau regions or to specify temporal 
fitting ranges.

The PDF for the complete set of fit parameters is given by
\begin{equation}
\label{EQN004} P(\Delta \mathcal{E}_{n,m} , c^j_{m,n}) = \frac{1}{Z} e^{-\frac{\chi^2}{2}}
\end{equation}
with appropriate normalization $Z$ and
\begin{equation}
\chi^2 = \sum_{j,k} \sum_{t/a=1}^{(T-1)/a} \frac{(C_{j k}^{\textrm{lattice}}(t)- C_{j k}(t))^2}{(w_{j k}(t))^2} , 
\end{equation}
which is the well-known $\chi^2$ used in $\chi^2$ minimizing fits ($C_{j k}^{\textrm{lattice}}(t)$ denote lattice
QCD results for correlators with corresponding statistical errors $w_{j k}(t)$). To obtain the PDF for a 
specific fit parameter, one has to integrate in (\ref{EQN004}) over all other parameters. In particular, the
probabilty for parameter $\mathcal{A}_j$ to be inside $[a,b]$ ($\mathcal{A}_j$ represents either one of the energy 
eigenvalue differences $\Delta \mathcal{E}_{n,m}$ or amplitudes $c^j_{m,n}$) is
\begin{equation}
\Pi(\mathcal{A}_j \in [a,b]) = \frac{
          \int_a^b d\mathcal{A}_j \,
          \int_{-\infty}^{+\infty} \prod_{k \ne j} d\mathcal{A}_k \,
           e^{-\chi^2/2}
          }
          {
          \int_{-\infty}^{+\infty} \prod_k d\mathcal{A}_k \,
          e^{-\chi^2/2}
          } .  
\end{equation}
This multi-dimensional integral can be computed with standard Monte Carlo methods. We implemented 
a parallel-tempering scheme combined with the Metropolis algorithm as described in detail in 
Ref.~\cite{Alexandrou:2014mka}. The parallel-tempering scheme prevents that the algorithm is stuck in a 
region around a local minimum of $\chi^2$ and, thus, guarantees ergodicity of the algorithm.

\begin{figure}[htb]
\begin{center}
\begin{minipage}[c]{.49\textwidth}
  \begin{center}
  \vspace{0.15cm}
  
  \includegraphics*[angle=0,width=\textwidth]{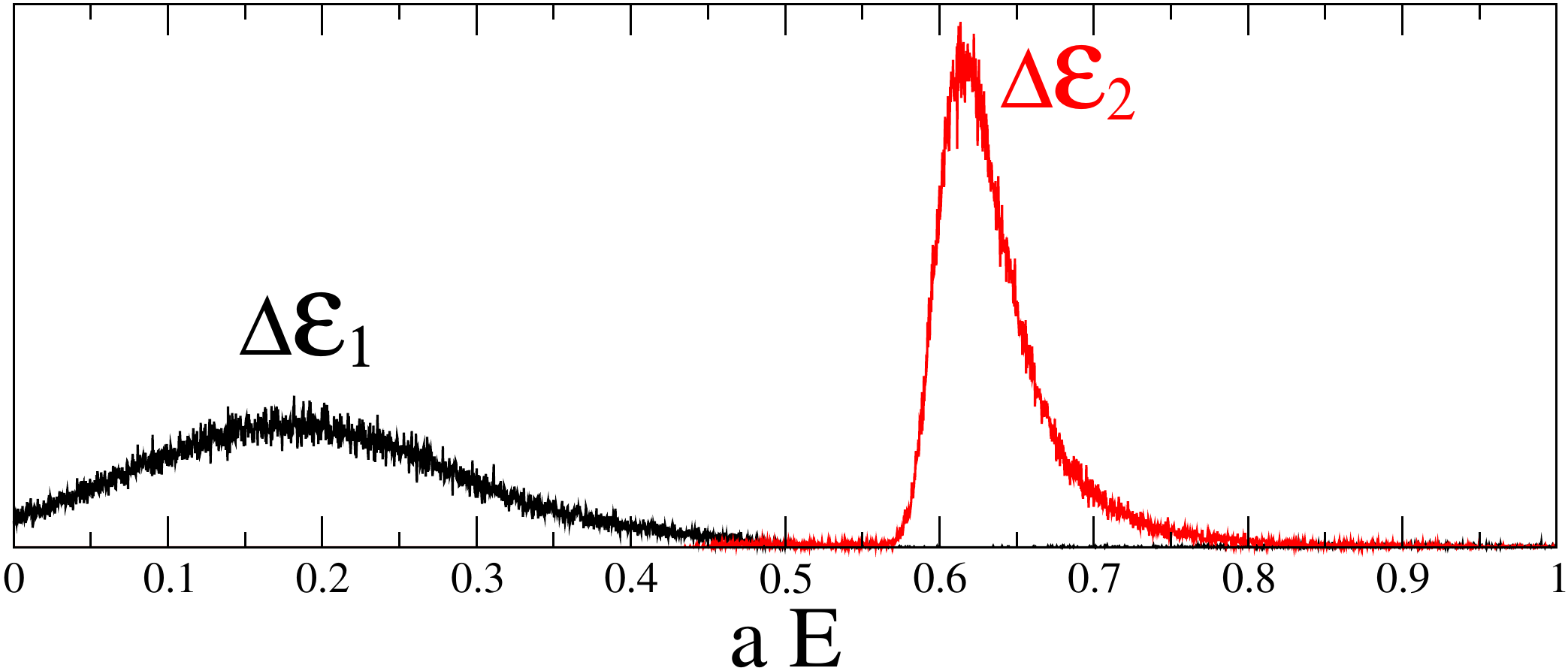} 
  \end{center}
\end{minipage}
\begin{minipage}[c]{.49\textwidth}
  \begin{center}

  \includegraphics*[angle=0,width=\textwidth]{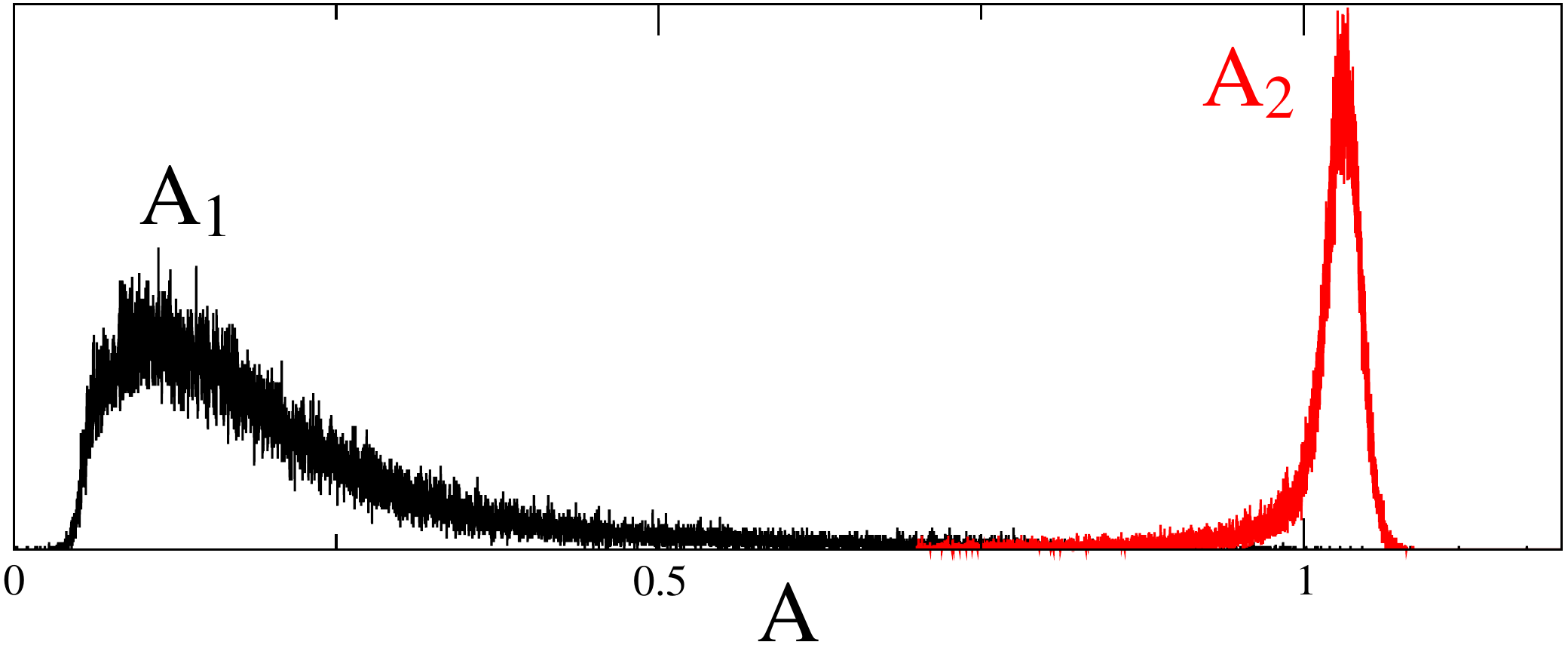}
  \end{center}

\end{minipage}
\caption{AMIAS analysis of the correlator of $\mathcal{O}^{q \bar{q}}$. PDFs for the parameters $\Delta \mathcal{E}_1$, $\Delta \mathcal{E}_2$, $A_1$ and $A_2$ from (\protect\ref{EQN005}).}
\label{fig:qqbarAMIAS}
\end{center}
\end{figure}

We have again analyzed the correlator of $\mathcal{O}^{q \bar{q}}$ shown in Fig.~\ref{fig:qqbarcor} (left), this time using AMIAS with fit function
\begin{equation}
\label{EQN005} C(t) = \sum_{n=1}^2 A_n \textrm{cosh}\{\Delta \mathcal{E}_n (t - T/2)\} .
\end{equation}
The resulting PDFs for the four parameters $\Delta \mathcal{E}_1$, $\Delta \mathcal{E}_2$, $A_1$ and $A_2$ are shown in Fig.~\ref{fig:qqbarAMIAS}.
As in Fig.~\ref{fig:qqbarcor} (right) two energy differences can clearly be identified, $a \Delta \mathcal{E}_1 \approx 0.25$ 
(the mass difference $m_\eta - m_\pi$) and $a \Delta \mathcal{E}_2 \approx 0.60$ (the mass of a $\pi + \eta$ or $K + \overline{K}$ two-meson state).

\vspace{-0.3cm}

\vspace{-1.4cm}
\section{Analysis of the $6 \times 6$ correlation matrix}
\label{sec:cormat}

We now use AMIAS to analyze the $6 \times 6$ correlation matrix with interpolating fields (\ref{EQN002}) to (\ref{EQN003}).
The fit function is obtained by restricting (\ref{eq:cosh}) to a finite number of terms,
\begin{equation}
\label{EQN006} C_{j k}(t) = \sum_{n=1}^N c^j_n (c^k_n)^\ast \textrm{cosh}\{\Delta \mathcal{E}_n (t - T/2)\} .
\end{equation}

\begin{figure}[htb]
\begin{center}
\vspace{-0.35cm}

\includegraphics*[angle=0,width=0.75\textwidth]{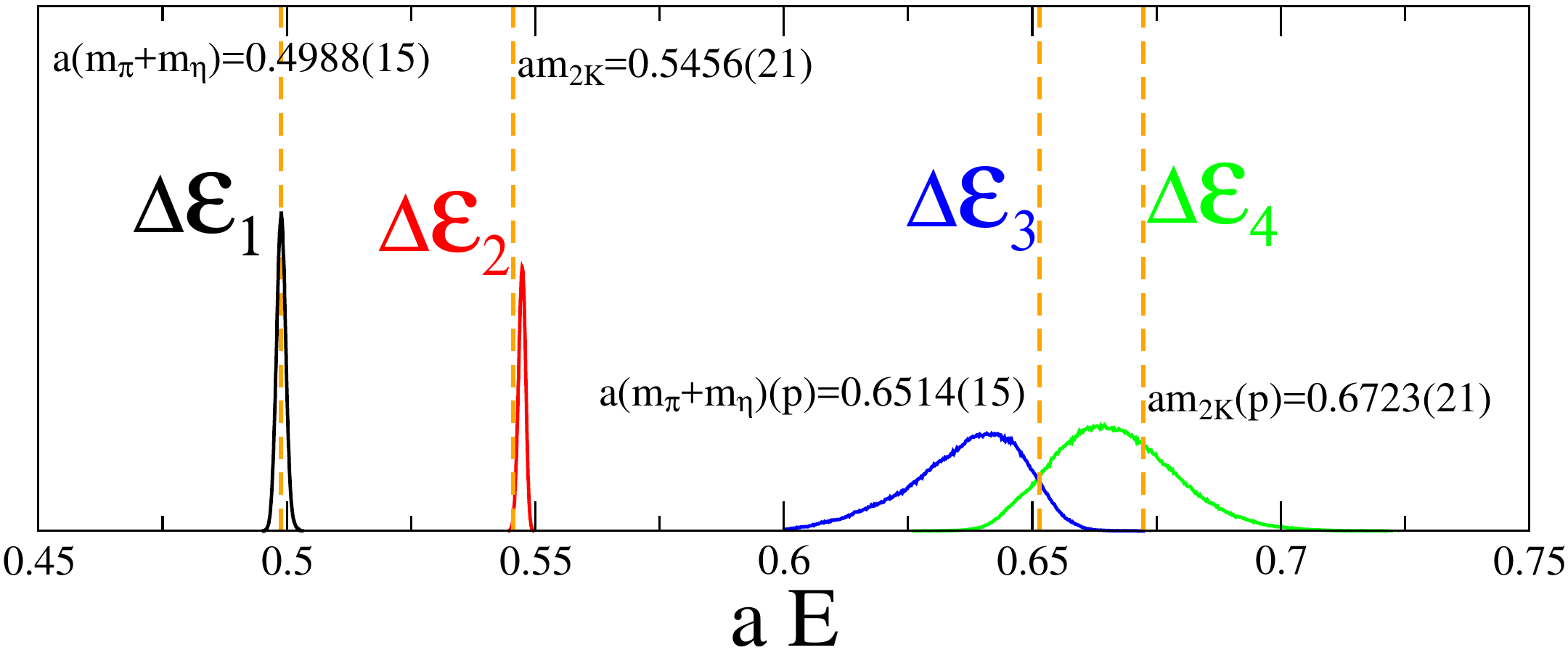}
\includegraphics*[angle=0,width=0.85\textwidth]{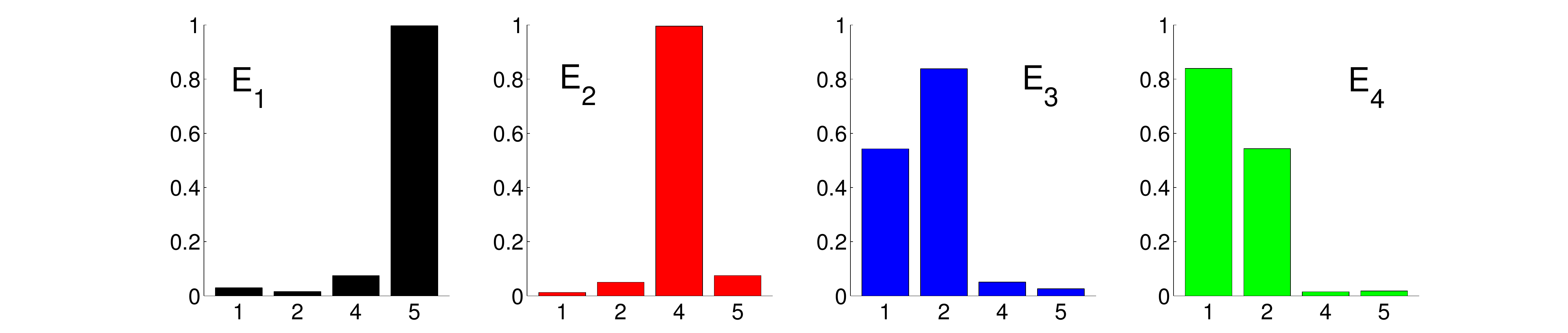}
\caption{AMIAS analysis of the $4 \times 4$ correlation matrix with interpolating fields 
$\mathcal{O}^2 = \mathcal{O}^{K \bar{K}, \ \textrm{point}}$, $\mathcal{O}^3 = \mathcal{O}^{\eta_{s} \pi, \ \textrm{point}}$, $\mathcal{O}^5 = \mathcal{O}^{K \bar{K}, \ \textrm{2part}}$ and 
$\mathcal{O}^6 = \mathcal{O}^{\eta_{s} \pi, \ \textrm{2part}}$ (propagation of strange 
quarks within the same timeslice neglected). Top: PDFs for the parameters
$\Delta \mathcal{E}_2 , \ldots , \Delta \mathcal{E}_5$ from (\protect\ref{EQN006}).
Bottom: $|v^j_n|$, the corresponding overlaps to the trial states $\mathcal{O}^{j \dagger} | \Omega \rangle$.}

\vspace{-0.6cm}
\label{fig:AMIAS}
\end{center}
\end{figure}

In a first analysis we use lattice QCD data, where the propagation of strange quarks within the same 
timeslice has been neglected. The interpolating field
$\mathcal{O}^{q \bar{q}}$ then decouples and, hence, cannot be considered in a tetraquark study 
of $a_0(980)$ (cf.\ also Refs.\ \cite{Abdel-Rehim:2014zwa,Berlin:2015faa}, where the same data has been used). 
Moreover, we do not consider the diquark-antidiquark interpolating field $\mathcal{O}^{Q \bar{Q}}$. In 
Fig.~\ref{fig:AMIAS} (top) we show the four lowest masses obtained with $N = 8$ terms in 
(\ref{EQN006})\footnote{AMIAS is able to determine the optimal $N$ automatically. For details we refer to \cite{Alexandrou:2014mka}.}, 
corresponding to the expected two-particle states $\pi + \eta$ and $K + \overline{K}$ with both mesons at rest
($\Delta \mathcal{E}_2$ and $\Delta \mathcal{E}_3$) as well as with one quantum of relative momentum ($\Delta \mathcal{E}_4$ and $\Delta \mathcal{E}_5$).
Note that AMIAS also finds the energy eigenvalue difference $m_\eta - m_\pi$ ($\Delta \mathcal{E}_1$, not shown in Fig.~\ref{fig:AMIAS}),
which has already been discussed in previous sections.

Since $(c^j_n)^\ast = \langle n | \mathcal{O}^{j \dagger} | \Omega \rangle$, the amplitudes extracted with AMIAS are the coefficients of 
the expansions of the trial states $\mathcal{O}^{j \dagger} | \Omega \rangle$ in terms of the extracted energy eigenstates $| n \rangle$, i.e.\
\begin{equation}
\mathcal{O}^{j \dagger} | \Omega \rangle \approx \sum_{n=1}^N (c^j_n)^\ast | n \rangle .
\end{equation}
More interesting, however, are the opposite expansions, i.e.\ the extracted energy eigenstates in terms of the trial states,
\begin{equation}
| n \rangle \approx \sum_{j=2,3,5,6} v^j_n \mathcal{O}^{j \dagger} | \Omega \rangle .
\end{equation}
It is easy to show that the matrix formed by the coefficients $v^i_n$ is the inverse of the matrix formed by the coefficients $(c^j_n)^\ast$ ($j = 2,3,5,6$, $n = 2,\ldots,5$) up to 
exponentially small corrections, i.e.\ $\sum_j v^j_m (c^j_n)^\ast = \delta_{m,n}$. Alternatively, 
one can also use the AMIAS samples for $\Delta \mathcal{E}_n$ and $c^j_n$ and Eq.~(\ref{EQN006}) to reconstruct a correlation matrix for each sample. Solving standard GEVPs for each reconstructed 
correlation matrix yields eigenvectors with components identical to $v^j_n$. The coefficients $|v^j_n|$ are shown in Fig.~\ref{fig:AMIAS} (bottom) (not the PDFs, but the most likely values).
Clearly, $\Delta \mathcal{E}_2$ and $\Delta \mathcal{E}_3$ correspond to two-particle states $\pi + \eta$ and $K + \overline{K}$ with 
both mesons at rest, since the coefficients $v^j_n$ show almost exclusively overlap to the trial states
$\mathcal{O}^{\eta_{s} \pi, \ \textrm{2part} \dagger} | \Omega \rangle$ and $\mathcal{O}^{K \bar{K}, \ \textrm{2part} \dagger} | \Omega \rangle$. $\Delta \mathcal{E}_4$ 
and $\Delta \mathcal{E}_5$ are close to the expected two-particle states with one quantum of relative momentum and the overlaps
shown by $v^j_n$ are consistent with this interpretation. There is no indication for any additional state in the region of $1000 \, \textrm{MeV}$, 
which could correspond to $a_0(980)$. These findings are consistent with our previous study using ETMC gauge link configurations \cite{Alexandrou:2012rm}.

\begin{figure}[htb]
\begin{center}
\includegraphics*[angle=0,width=0.75\textwidth]{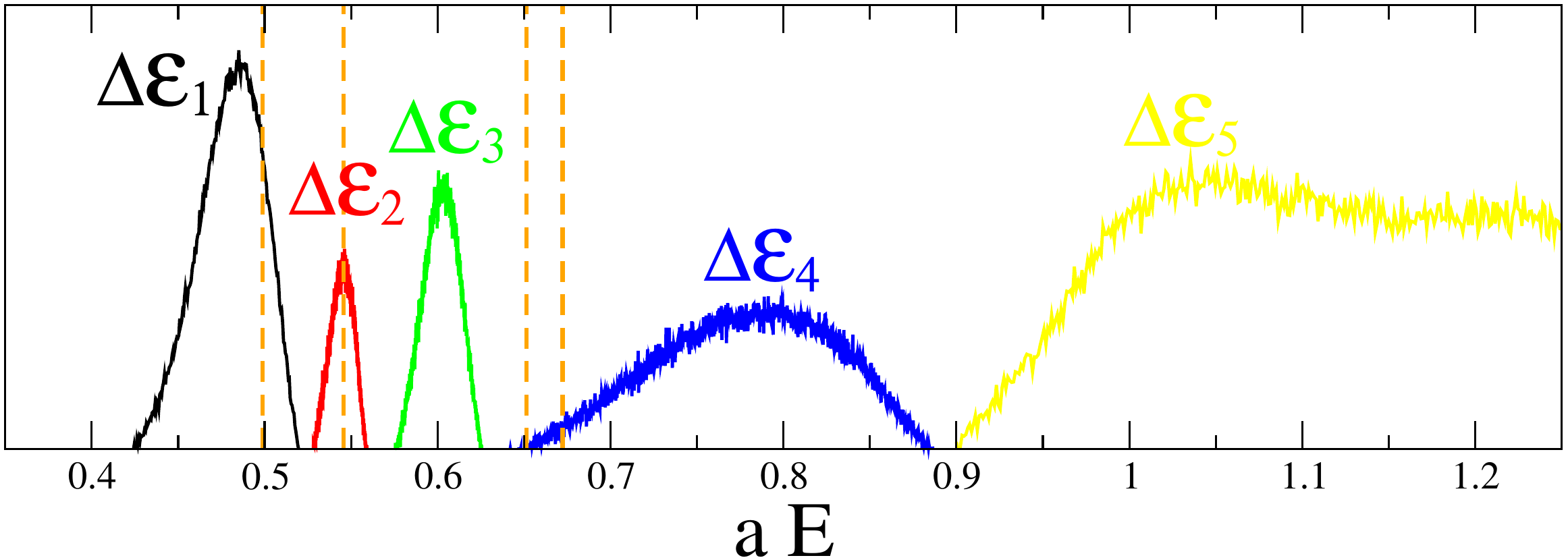}
\vspace{-0.2cm}
\caption{AMIAS analysis of the full $6 \times 6$ correlation matrix (propagation of 
strange quarks within the same timeslice taken into account). PDFs for the parameters
$\Delta \mathcal{E}_2 , \ldots , \Delta \mathcal{E}_6$ from (\protect\ref{EQN006}) generated by AMIAS.}
 \label{fig:AMIASwl}
\end{center}
\end{figure}

When including the propagation of strange quarks within the same timeslice, the interpolating field $\mathcal{O}^{q\bar{q}}$ couples 
to the four-quark interpolating fields given in Eqs.~(\ref{EQN007}) to (\ref{EQN003}). This introduces, however, a lot of statistical noise and 
it is rather difficult to resolve energy differences precisely. In Fig.~\ref{fig:AMIASwl} we show preliminary AMIAS results for
the full $6 \times 6$ correlation matrix, where $N = 12$ terms in Eq.~(\ref{EQN006}) have been used. The two lowest energy eigenstates exhibit
rather clear signals and correspond to the $\pi + \eta$ and $K + \overline{K}$ two-meson states. Even though higher excitations are less prominent,
we find strong indication for an additional state with mass $\Delta \mathcal{E}_3 \approx 0.6 / a \approx 1300 \, \textrm{MeV}$, which
is below the expectation for the two-particle states with one quantum of relative momentum and, hence, might be a candidate for the $a_0(980)$.

\section{Conclusions}

With AMIAS one can extract energy differences and amplitudes without the necessity to identify pleateau regions or to specify temporal 
fitting ranges. We have shown that AMIAS can successfully be used to analyze correlators with strong contributions from several energy 
eigenstates as well as rather noisy correlation matrices. An example for the latter is e.g.\ the full $6 \times 6$ correlation matrix 
for the $a_0(980)$ channel. Another major advantage of AMAIAS might be, that it does not require all elements of a correlation matrix, 
i.e.\ particularly noisy elements or elements, which are very time-consuming to compute, can be omitted. This we plan to address in a future publication.



\noindent\textbf{Acknowledgements:}
M.W.\ acknowledges support by the Emmy Noether Programme of the DFG (German Research Foundation), 
grant WA 3000/1-1. This work was supported in part by the Helmholtz International Center for FAIR within the framework 
of the LOEWE program launched by the State of Hesse.
This work was cofunded by the European Regional Development Fund and the Republic of Cyprus through the Research
Promotion Foundation (Project Cy-Tera NEA $\textrm{Y}\Pi\textrm{O}\Delta\textrm{OMH}$/$\Sigma\textrm{TPATH}$/$0308$/$31$) by the grand cypro914.



\end{document}